\documentclass[]{spie}

\usepackage{amsmath,graphicx,tabularx,ulem}

\title{Measurements of binary stars with coherent integration of NPOI data}

\author{Anders M. Jorgensen\supit{a}, H. Schmitt\supit{b,f},
  R. Hindsley\supit{b}, J. T. Armstrong\supit{b},
  T. A. Pauls\supit{b},\\ D. Mozurkewich\supit{d},
  D. J. Hutter\supit{c}, C. Tycner\supit{e} \skiplinehalf \supit{a}New
  Mexico Institute of Mining and Technology, Socorro, NM,
  USA\\ \supit{b}Naval Research Laboratory, Washington, DC,
  USA\\ \supit{c}Naval Observatory Flagstaff Station, Flagstaff, AZ,
  USA\\ \supit{d}Seabrook Engineering, Seabrook, MD,
  USA\\ \supit{e}Central Michigan University, Mt. Pleasant, MI,
  USA\\ \supit{f}Interferometrics, Inc., Herndon, VA,
  USA\\ \footnotemark[0]}

\begin{document}

\twocolumn[{\csname @twocolumnfalse\endcsname
\vspace{0.2in}
\maketitle

\begin{abstract}
  In this paper we use coherently integrated visibilities (see
  separate paper in these proceedings\cite{jorgensen:2008}) to measure
  the properties of binary stars. We use only the phase of the complex
  visibility and not the amplitude. The reason for this is that
  amplitudes suffer from the calibration effect (the same for coherent
  and incoherent averages) and thus effectively provide lower accuracy
  measurements. We demonstrate that the baseline phase alone can be
  used to measure the separation, orientation and brightness ratio of
  a binary star, as a function of wavelength.
\end{abstract}
\vspace{0.1in}
}]

\section{Introduction}

\noindent Binary stars are important for calibrating evolutionary
stellar models. Because stellar models are sensitive to the parameters
of the stars, it is important to obtain the highest possible accuracy
of the stellar parameters. In this paper we will demonstrate how to
extract brightness ratio and separation vector from measurements of
the complex visibility phase only. This is significant because the
visibility phase does not suffer from the calibration effects of
visibility amplitudes, and therefore much higher precision can be
obtained. Further, there are more complex visibility baseline phases
than closure phases such that using baseline phases instead of closure
phases yields more information. Finally, complex visibility phases
generally have better SNR than closure phases.

\section{Theory}

\noindent The complex visibility of a binary star is 

\begin{equation*}
  \tilde{V}=\cos\left(r\gamma\right)+r\cos\left(\gamma\right)+i\left[\sin\left(r\gamma\right)+r\sin\left(\gamma\right)\right]
\end{equation*}

\noindent where

\begin{equation*}
  \gamma=\frac{2\pi\vec{B}\cdot\vec{s}}{\left(r+1\right)\lambda}.
\end{equation*}

\noindent $\vec{B}$ is the baseline vector, $\vec{s}$ is the
separation vector, $r$ is the brightness ratio, and $\lambda$ is the
wavelength. The visibility phase is

\begin{equation}
  \label{eq_binary_phase}
  \theta=\tan^{-1}\left(\frac{\sin\left(r\gamma\right)+r\sin\left(\gamma\right)}{\cos\left(r\gamma\right)+r\cos\left(\gamma\right)}\right)
\end{equation}

\noindent with uncertainty

\begin{equation*}
  \sigma_\theta=\frac{1}{\sqrt{2NV^2}}
\end{equation*}

\noindent where $N$ is the total number of photons counted and $V$ is
the coherently integrated visibility amplitude.

In addition to the source phase, the measured phase also contains an
instrumental phase component and a combination of atmospheric and
vacuum phase terms,

\begin{equation}
  \label{eq_total_phase}
  \theta=\theta_\text{source}+\theta_\text{inst}+\theta_\text{atm}
\end{equation}

\begin{figure}
\includegraphics[width=\linewidth]{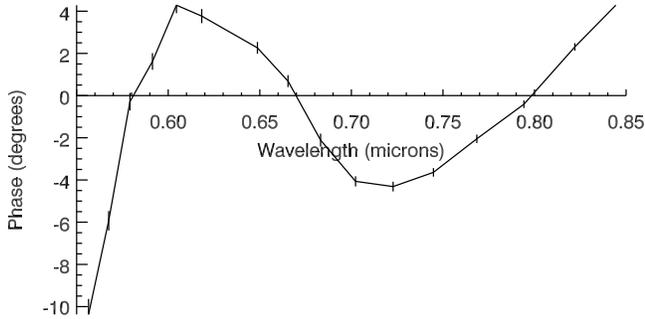}
\caption{\label{figure_phase_of_calibrator} The phase of a calibrator star. This is the sum of the instrumental phase and some atmospheric phase.}
\end{figure}

\noindent The instrumental phase term can be measured by observing a
calibrator star (which has zero source phase). The observed phase of a
calibrator may contain both instrumental and atmospheric phase terms,
which is not a problem because the phase terms are
additive. Figure~\ref{figure_phase_of_calibrator} shows the measured
phase of a calibrator star. The atmospheric phase term takes the form

\begin{equation}
  \label{eq_atm_phase}
  \theta_\text{atm}=\frac{2\pi}{\lambda}\left[v+\left(n-1\right)a\right]+\phi
\end{equation}

\noindent where $v$ is a vacuum path, $a$ is the atmosphere path, $n$
is the wavelength-dependent index of refraction, and $\phi$ is a phase
offset.

\begin{figure}
\includegraphics[width=\linewidth]{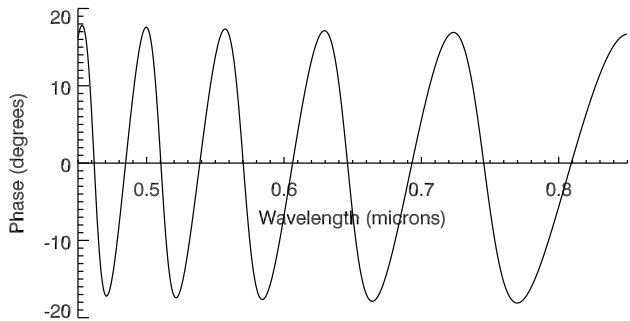}
\caption{\label{figure_binary_phase} Example of what a binary star
  baseline might look like in a fringe-tracking interferometer.}
\end{figure}

We can then measure the brightness ratio and separation vector by
fitting equation~\ref{eq_total_phase} to the baseline phases, with
equation~\ref{eq_binary_phase} inserted for the source phase, and
equation~\ref{eq_atm_phase} inserted for the atmospheric phase
term. Note that the binary phase term in
equation~\ref{eq_binary_phase} is monotonic as it stands, but in a
typical measurement we will see the phase be roughly centered around
zero. This is because the fringe-tracker of the interferometer will
add the appropriate vacuum delay to give the fringe close to zero
phase. Figure~\ref{figure_binary_phase} shows an example of what the
phase of a binary might realistically look like in at high spectral
resolution in a fringe-tracking instrument interferometer like the
NPOI\cite{armstrong:1998}.

\section{Data sets}

\noindent We will present measurements of two stars,
$\theta^2\,\text{Tauri}$ and $\kappa\,\text{Ursa Majoris}$ observed on
the same night. On that night the NPOI was configured to observe three
baselines, one of which was observed on two
spectrographs. Table~\ref{table_observations} lists the target
observations. In addition to target observations there were several
calibrator observations to measure the instrumental phase for
subtraction.

\begin{table}
\centerline{\begin{tabular}{lrrr}
\hline
Date & Time (UT) & Target & Hour angle \\
\hline
2004/1/30 & 04:43 & $\theta^2\,\text{Tau}$ & 1.373\\
2004/1/30 & 06:09 & $\theta^2\,\text{Tau}$ & 2.810\\
2004/1/30 & 07:15 & $\theta^2\,\text{Tau}$ & 3.923\\
\hline
2004/1/30 & 06:37 & $\kappa\,\text{UMa}$ & -1.284\\
2004/1/30 & 09:31 & $\kappa\,\text{UMa}$ & 1.612\\
2004/1/30 & 10;26 & $\kappa\,\text{UMa}$ & 2.499\\
2004/1/30 & 10:47 & $\kappa\,\text{UMa}$ & 2.871\\
2004/1/30 & 11:26 & $\kappa\,\text{UMa}$ & 3.541\\
\hline
\end{tabular}}
\caption{\label{table_observations} Observations of
  $\theta^2\,\text{Tauri}$. All observations lasted $30\,\text{s}$.}
\end{table}

\begin{figure}
\includegraphics[width=\linewidth]{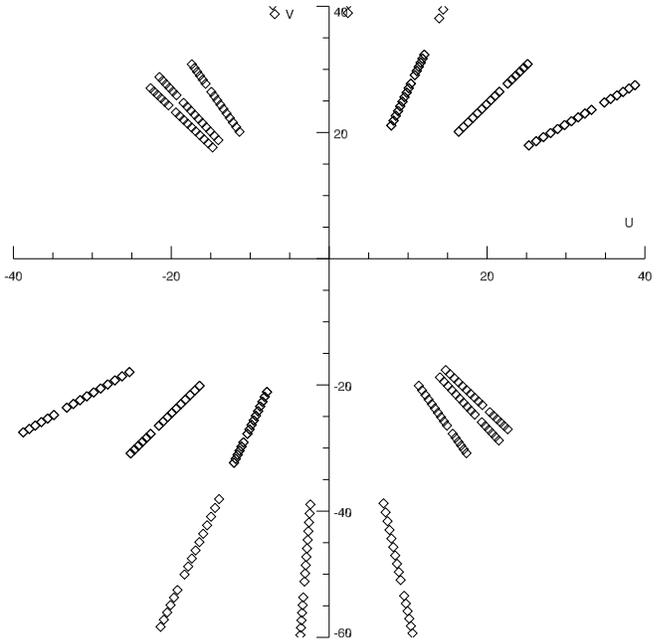}
\caption{\label{figure_tet2tauUvPlot} UV coverage on $\theta^2\,\text{Tauri}$.}
\end{figure}

\section{Analysis}

\noindent The coherently integrated visibilities are produced using
the methodology presented in the paper \cite{jorgensen:2008} in these
proceedings. The fit of the functional form of
equation~\ref{eq_total_phase} is not trivial. There are many false
minimas such that it is important to either use a minimization method
which finds the global minimum and ignores local minima, or to give
the minimization procedure an initial guess which is within the
capture region of the global minimum. In this paper we are taking the
second approach and use a grid-search to obtain an approximate initial
guess. If we examine
equations~\ref{eq_total_phase},~\ref{eq_atm_phase},
and~\ref{eq_binary_phase} we note that there are three parameters for
the binary, $r$ (brightness ratio), $\alpha$, and $\beta$ (separation
along the $u-$ and $v$-axes respectively). In addition, there are
three parameters for each baseline, $v$, $a$, and $\phi$. For $N$
baselines there are thus $3\times\left(N+1\right)$ parameters, and it
is not feasible to do a grid search over that many
parameters. Fortunately, the coherent integration process results in
an average phase which is close to zero. If we therefore use a binary
phase model which has approximately zero mean phase (we achieve this
by adjusting the vacuum path) then we can in many cases reduce the
grid search to cover just the parameters $r$, $\alpha$, and
$\beta$. If we make a reasonable guess for $r$ from the amplitude of
the phase variations then the grid search is reduced to just two
parameters.

\subsection{$\theta^2\,\text{Tauri}$}

\begin{figure}
\includegraphics[width=\linewidth]{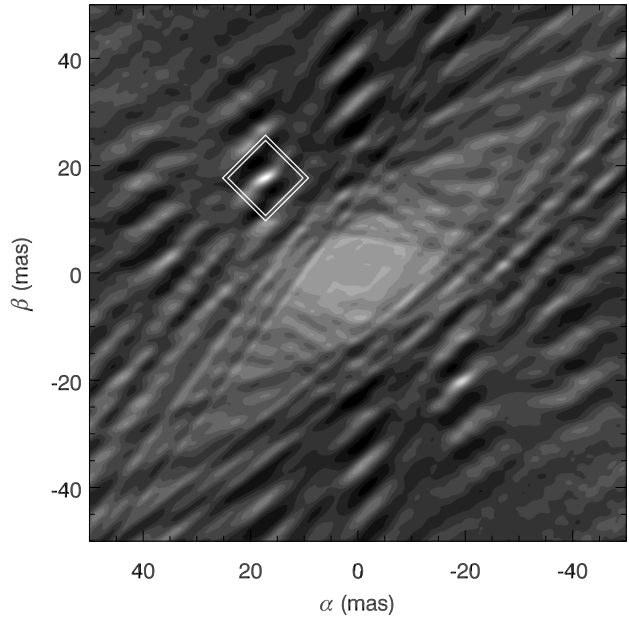}
\caption{\label{figure_tet2tauGridSearch} Result of grid search over
  $\alpha$, $\beta$ on $\theta^2\,\text{Tauri}$.}
\end{figure}

\noindent In the analysis of $\theta^2\,\text{Tauri}$ we begin with a
grid search as outlined above. Figure~\ref{figure_tet2tauGridSearch}
shows the goodness of fit as a function of the two separation
parameters. We chose a brightness ratio of $r=0.35$ based on examining
the amplitude of the phase oscillations in the raw data. The grayscale
of the figure indicates the goodness of fit according to a $\chi^2$
metric, with a brighter color indicating a better fit and a darker
color indicating a worse fit. The best fit, at $\alpha\approx
20\,\text{mas}$, and $\beta\approx 20\,\text{mas}$ is framed by two
white diamonds. Notice that there is also a good fit (but not quite as
good) at approximately $(-\alpha,-\beta)$. Finally, notice the
complexity of the fitness landscape. Without a good initial guess or a
optimization method which is able to ignore local minima it would be
unlikely that the fit converges on the global minimum.

\begin{figure*}
\includegraphics[width=\linewidth]{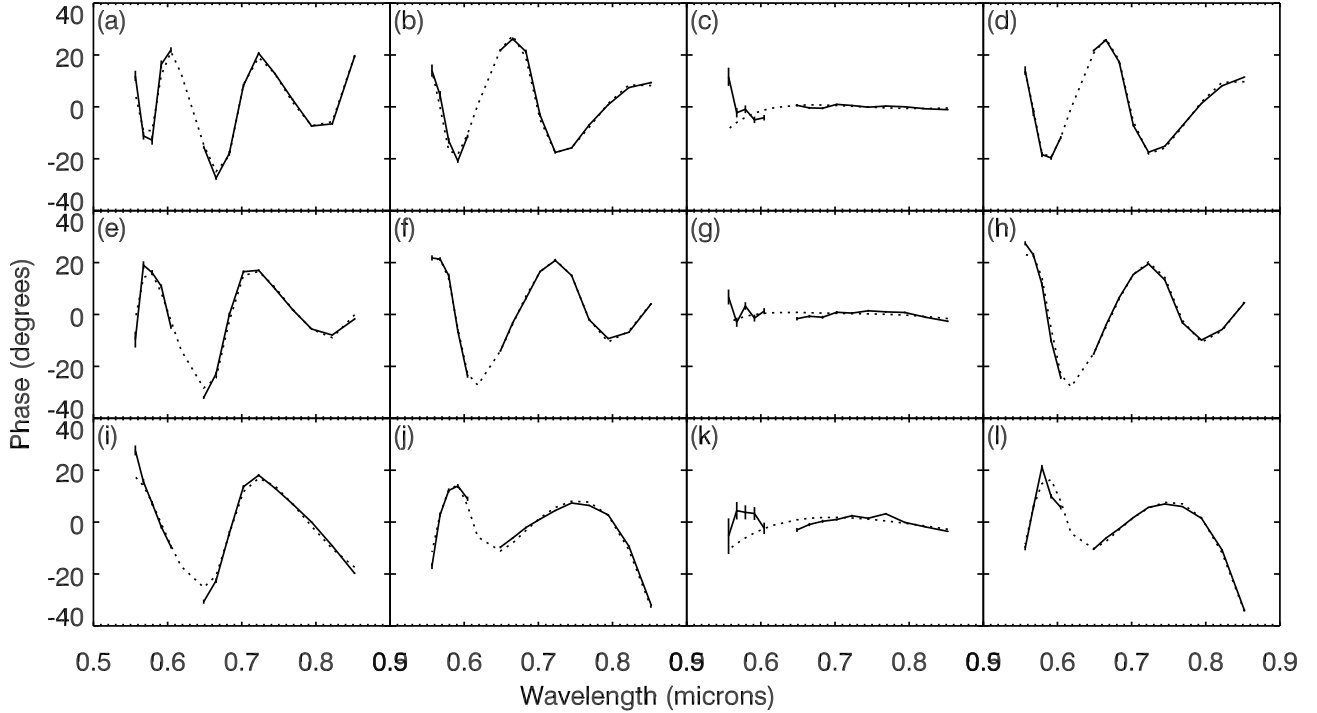}
\caption{\label{figure_tet2tauFit} Visibility phase (solid) and model
  (dotted) for $\theta^2\,\text{Tauri}$.}
\end{figure*}

Using this initial guess we then fit the full function
(Equation~\ref{eq_total_phase} with equations~\ref{eq_binary_phase},
and~\ref{eq_atm_phase} inserted) to the 12 baseline measurements
simultaneously (i.e. using the same binary star parameters for every
baseline). However, we make one modification in that we allow $r$ to
vary linearly with wavelength. We then obtain the fit shown in
figure~\ref{figure_tet2tauFit}. In this figure the solid curve shows
the baseline phase (including error bars which are often too small to
distinguish), and the dotted curve shows the best-fit model.

\begin{figure}
\includegraphics[width=\linewidth]{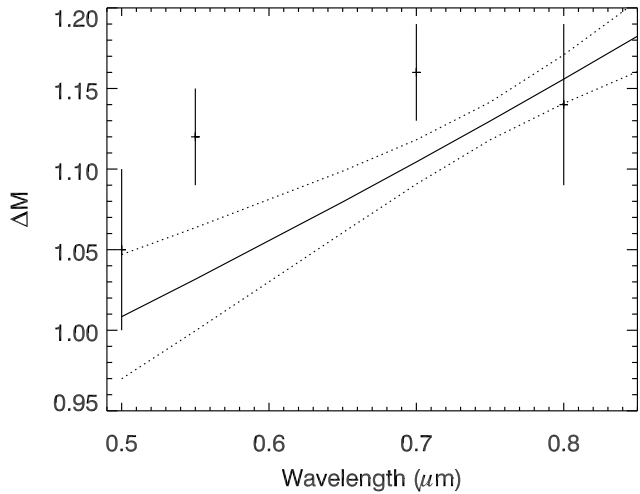}
\caption{\label{figure_brightness_ratio_vs_lambda} Brightness ratio as
  a function of wavelength (solid curve) and one standard deviation
  (dotted curves). Also plotted are individual data points taken from
  \cite{armstrong:2006}.}
\end{figure}

In order to estimate the uncertainty on the fitted parameters, $r$,
$\frac{dr}{d\lambda}$, $\alpha$, and $\beta$, we perform a bootstrap
Monte Carlo analysis. Bootstrap error analysis is a simple way of
estimating the sampling uncertainty from a data set with the same
distribution as the measured data set. It proceeds by repeatedly
selecting $N$ (with replacement) from the $N$ independent segment of
the data set and performing the full analysis to obtain parameters,
each time obtaining a slightly different set of parameters. If the
data set is a good representation of the distribution from which it is
taken (this is the usual assumption) then the distribution of
parameters represents the uncertainty of such a sample. We took 100
bootstrap samples and arrived at separation parameters of
$\alpha=17.134\pm0.033\,\text{mas}$,
$\beta=17.989\pm0.022\,\text{mas}$, which yields a precision on the
separation of approximately $0.16\%$. The brightness ratio as a
function of wavelength is plotted in
Figure~\ref{figure_brightness_ratio_vs_lambda}. In that figure the
solid curve is the brightness ratio, whereas the dotted curves
represent one standard deviation. Also plotted are measurements of the
brightness ratio from Armstrong et al. (2006)
\cite{armstrong:2006}. That paper used a much larger database than the
present work, spanning both Mark III and NPOI data, yet arrived at
uncertainties in the brightness ratio which are similar to or greater
than ours. Table~\ref{table_brightness_ratio} lists the brightness
ratio at several wavelengths. The discrepancy between this work and
prior work will elaborated on in the discussion section. For now we
just note that the reduced $\chi^2$ of the fit is approximately eight,
corresponding to a mean difference between model and data of 2.5
standard deviations.

\begin{table}
\begin{tabular}{rrr}
\hline
$\lambda\,(\mu\text{m})$ & This work & Armstrong et al. (2006)\\
\hline
0.50 & & $1.05\pm 0.05$\\
0.55 & $1.03\pm0.03$ & $1.12\pm 0.03$\\
0.65 & $1.08\pm 0.02$ & \\
0.70 & $1.10\pm 0.01$ & $1.16\pm 0.03$\\
0.75 & $1.13\pm 0.01$ & \\
0.80 & $1.16\pm 0.01$ & $1.14\pm 0.05$\\ 
\hline
\end{tabular}
\caption{\label{table_brightness_ratio} Brightness ratio as a function of wavelength for this work compared with Armstrong et al. (2006)}
\end{table}

\subsection{$\kappa\,\text{Ursa Majoris}$}

\begin{figure*}
\includegraphics[width=\linewidth]{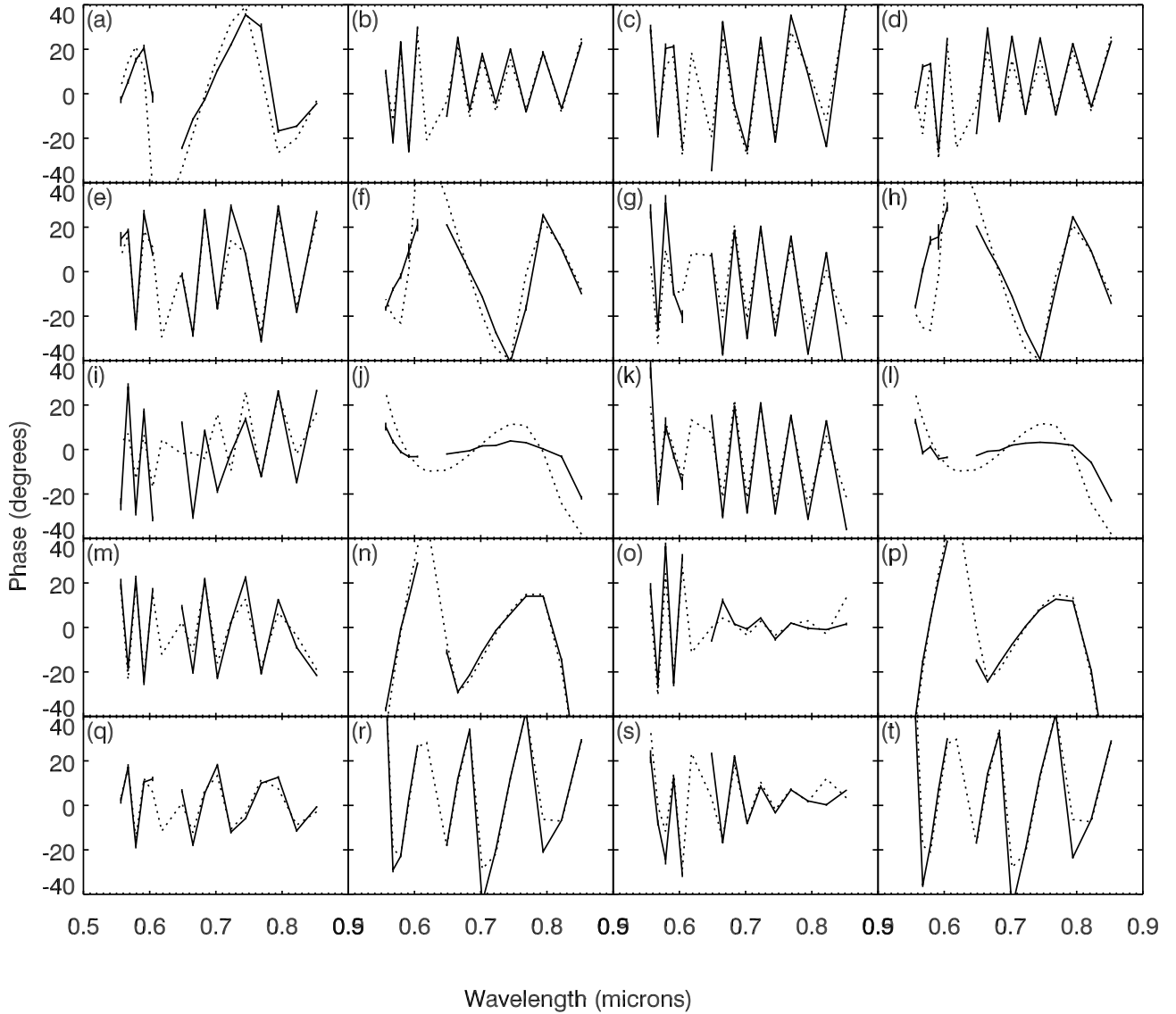}
\caption{\label{figure_kappaumaFit} Visibility phase (solid) and model (dotted) for $\kappa\,\text{Ursa Majoris}$.}
\end{figure*}

\noindent The second example, $\kappa\,\text{Ursa Majoris}$ is more
challenging. It consists of five observations on the same set of
baselines. We approach this data set in the same way as the previous
data set, and the final fit to the data is presented in
Figure~\ref{figure_kappaumaFit}. The fit corresponds to a brightness
ratio of $\Delta M\approx 0.5$, and separation
$\alpha=-115\,\text{mas}$, $\beta=85\,\text{mas}$. Because of
discrepancies between the model and the data we do not specify
uncertainties but instead refer to the discussion section.

\section{Discussion}

\noindent An important consideration when doing high-accuracy work is
knowledge of the wavelengths to a high degree of accuracy. In the case
of $\theta^2\,\text{Tauri}$ wavelength knowledge is less critical for
estimating the brightness ratio, but still important for the
separation. In the case of a large-separation binary system such as
$\kappa\,\text{Ursa Major}$ it is much more important. For both the
data sets analyzed here only nominal wavelength and bandpass
information was available, and that likely is a contributing factor to
the imperfect fit $\chi^2=8$ in the case of $\theta^2\,\text{Tauri}$
and is probably the dominant effect of the goodness of the fit in the
case of $\kappa\,\text{Ursa Major}$. In
Figure~\ref{figure_kappaumaFit}, panels (b) and (d) contain the same
baseline, but recorded on two different spectrographs. However the two
observations are not identical, which is especially evident at the
blue end of the spectrum. The likely cause of this is that the two
spectrographs have slightly different wavelength scales. We are not
able to correct this effect because no measured wavelength information
was available for that particular date. However on later dates the
bandpass is measured to a high degree of accuracy.

Another important thing to notice, if we look at
Figure~\ref{figure_brightness_ratio_vs_lambda}, is the discrepancy
between the present results and those of Armstrong et
al. (2006)\cite{armstrong:2006}. First, in the older data we observe
what appears to be a curvature as a function of wavelength. We do not
see this in our data because we modeled the brightness ratio to be
linear with wavelength. It would be interesting to add a quadratic
term and then repeat the comparison. Another thing to note is that
Armstrong et al. (2006) obtain slightly larger brightness ratios than
us. It is important to realize in this context that the amplitude of
the phase oscillation is modulated not only by the brightness ratio,
but also by the visibility amplitude of the component stars. The
component stars are unresolved such that visibilities are close to
one. We did not take that effect into account in this initial work.

\section{Conclusion}

\noindent We have demonstrated that coherently integrated visibility
phases can be used to obtain high-quality measurements of the
fundamental parameters of binary stars. The measurements are accurate
to the point where careful calibration of wavelengths and bandpasses
is necessary. This calibration was not available for the data sets
considered in this paper, but is routinely performed on more recent
NPOI data sets.

\acknowledgments

The NPOI is funded by the Office of Naval Research and the
Oceanographer of the Navy. The authors would like to thank R. Zavala
for useful suggestions and for help with access to the relevant data.

\bibliographystyle{spiebib}
\bibliography{main}

\end{document}